\documentclass[a4paper,12pt]{article}
\usepackage[utf8]{inputenc}
\usepackage{amsmath,amsfonts,amsthm,amssymb,dsfont}
\usepackage{graphicx,wrapfig,lipsum}
\usepackage[section]{placeins}

\usepackage[numbers,sort&compress]{natbib}

\usepackage{tikz}
\usetikzlibrary{shapes.geometric,decorations.markings}
\usepackage{subfigure}

\usepackage[hidelinks]{hyperref}

\hypersetup{
    colorlinks=true,
    linktocpage=true,
    linkcolor=blue,
    urlcolor=blue,
    citecolor=blue,
}

\numberwithin{equation}{section}
\newcommand{\beq}{\begin{equation}}
\newcommand{\eeq}{\end{equation}}
\newcommand{\bea}{\begin{eqnarray}}
\newcommand{\eea}{\end{eqnarray}}
\newcommand{\nn}{\nonumber}

\newcommand{\be}{\begin{equation}}
\newcommand{\ee}{\end{equation}}

\usepackage{soul}

\definecolor{mycolor}{RGB}{255,238,140}

\newcommand{\calA}{\mathcal{A}}
\newcommand{\calD}{\mathcal{D}}
\newcommand{\calF}{\mathcal{F}}
\newcommand{\calN}{\mathcal{N}}
\newcommand{\calR}{\mathcal{R}}
\newcommand{\Nf}{N_{f}}
\newcommand{\Nc}{N_{c}}
\newcommand{\Mink}[1]{\text{Mink}_{#1}}
\newcommand{\tk}{\tilde{k}}

\begin{document}

\begin{titlepage}

\begin{center}

$\phantom{.}$\\ \vspace{2cm}
\noindent{\Large{\textbf{Twisted-Circle Compactifications of SQCD-like Theories and Holography
}}}

\vspace{1cm}

Niall T. Macpherson$^{\text{a,b}}$ \footnote{macphersonniall@uniovi.es}, Paul Merrikin$^{\text{c}}$ \footnote{2043506@swansea.ac.uk}, Carlos Nunez$^{\text{c}}$ \footnote{c.nunez@swansea.ac.uk}, Ricardo Stuardo$^{\text{a,b}}$ \footnote{ricardostuardotroncoso@gmail.com}

\vspace{0.5cm}

$^\text{a}$Departamento de Física, Universidad de Oviedo,\\
Avda. Federico García Lorca 18, 33007 Oviedo, Spain\\
\vspace{0.5cm}
$^\text{b}$Instituto Universitario de Ciencias y Tecnologías Espaciales de Asturias (ICTEA),\\ Calle de la Independencia 13, 33004 Oviedo, Spain\\
\vspace{0.5cm}
$^{\text{c}}$Department of Physics, Swansea University,\\ Swansea SA2 8PP, United Kingdom

\end{center}

\vspace{0.5cm}
\centerline{\textbf{Abstract}} 

\vspace{0.5cm}

\noindent{We construct and analyse holographic duals to a class of four-dimensional $\mathcal{N}=1$ SU$(N_c)$ SQCD-like theories compactified on a circle with an R-symmetry twist. The setup originates from type IIB backgrounds previously proposed as duals to SQCD with $\Nf$ fundamental flavours. The U(1) R-symmetry is anomaly-free only if $N_f = 2N_c$. We implement a supersymmetric twisted-circle reduction, holographically realised through a smoothly shrinking S$^1$ fibered over the internal U(1)$_\text{R}$ direction. We obtain new regular type IIB supergravity backgrounds that are valid only if the condition $N_f = 2N_c$ is satisfied—mirroring the anomaly cancellation requirement in the field theory.
%
We compute various field-theoretic observables—including the Chern-Simons level, Wilson loop and the holographic central charge—showing the emergence of a 3D $\mathcal{N}=2$ gapped phase consistent with a Chern-Simons TQFT. This work highlights the interplay between anomalies, supersymmetry, and geometry in the holographic realisation of compactified gauge theories with fundamental matter.}

\vspace*{\fill}

\end{titlepage}

\newpage

\tableofcontents
\thispagestyle{empty}

\newpage
\setcounter{page}{1}
\setcounter{footnote}{0}

\section{Introduction}
After Maldacena's conjecture of the AdS/CFT duality \cite{Maldacena:1997re}, subsequent refinements \cite{Gubser:1998bc, Witten:1998qj} and extension to non-conformal theories \cite{Itzhaki:1998dd, Boonstra:1998mp}, it became natural to apply these ideas to study non-perturbative aspects of generic Quantum Field Theories (QFTs). This led to the construction of supergravity backgrounds holographically dual to QFTs with multiple dynamical scales and interesting non-perturbative phenomena such as confinement and symmetry breaking. Examples include \cite{Witten:1998zw, Girardello:1999bd, Klebanov:2000hb, Maldacena:2000yy}, among others.

More recently, progress has been made using models based on wrapped branes. In particular, string backgrounds have been studied that correspond to the compactification of a $d$-dimensional supersymmetric field theory on a fixed-size twisted-circle S$^{1}$, yielding a gapped QFT on $\mathbb{R}^{1,(d-2)} \times \text{S}^1$. At low energies, excitations propagate in $(d-1)$ dimensions, while at higher energies the circle opens up and conformal dynamics is recovered. 
See \cite{Anabalon:2021tua,Bobev:2020pjk,Anabalon:2022aig,Nunez:2023nnl,Nunez:2023xgl, Fatemiabhari:2024aua,Chatzis:2024top,Chatzis:2024kdu,Anabalon:2024che,Castellani:2024pmx,Fatemiabhari:2024lct, Kumar:2024pcz, Castellani:2024ial, Chatzis:2025dnu} for some recent works.

Of particular interest here is the compactification of 4D theories on a twisted-circle of radius $R_0$, as explored in \cite{Cassani:2021fyv}. For any SU$(\Nc)$ supersymmetric gauge theory with a \textit{non-anomalous} U(1) R-symmetry, it is possible to couple the theory to a constant background gauge field along the S$^1$. When combined with anti-periodic boundary conditions for fermions, this setup restores a massless 3D ${\cal N}=2$ vector multiplet, while the 4D chiral multiplets generate a KK mass spectrum that is not vector-like. At energies below $\frac{1}{R_0}$, the KK tower of massive chiral multiplets can be integrated out, resulting in a Chern-Simons level $k = \Nc$. Crucially, this relies on the non-anomalous nature of the U(1) R-symmetry and holds for any number of chiral multiplets in any representation of the SU($\Nc$) gauge group.

In this paper, we focus on a 4D SQCD-like theory obtained by reducing 6D ${\cal N}=(1,1)$ SU$(\Nc)$ SYM on a sphere with a twist, and then coupling it to $\Nf$ massless chiral multiplets in the fundamental representation of the gauge group. The compactification yields a 4D ${\cal N}=1$ SU($\Nc$) theory coupled to an infinite tower of KK modes (massive vector and chiral multiplets) and to fundamental flavours. These theories were studied holographically in \cite{Casero:2006pt, Casero:2007jj, Hoyos-Badajoz:2008znk, Caceres:2007mu, Nunez:2010sf}, where it was shown that the U(1) R-symmetry is non-anomalous only when $\Nf = 2\Nc$. Hence, further compactification on a twisted circle is viable only in this case.
\subsection{General idea of this paper}
As stated above, we work with a dual pair consisting of a Type IIB background and a QFT, originally studied in \cite{Casero:2006pt, Casero:2007jj, Hoyos-Badajoz:2008znk}. The field theory, a deformation of 4D ${\cal N}=1$ SQCD, preserves a U$(1)$ R-symmetry only when $N_f = 2N_c$. In this work, we consider the twisted-circle compactification of this specific version of SQCD. The holographic dual is constructed using the generating technique developed in \cite{Macpherson:2024qfi}.

The dual backgrounds exhibit a smoothly shrinking S$^1$ which, together with the radial coordinate, forms a cigar-like geometry, implementing the compactification of the field theory on S$^1$. Additionally, the shrinking S$^1$ is fibered over an internal U(1) direction associated with the R-symmetry, realizing the twist between the R-symmetry and the spacetime U(1).

These geometries are systematically constructed following \cite{Macpherson:2024qfi}. In compactifications of the form $\Mink{D+1} \rightarrow \Mink{D}$ in Type II theories, the supersymmetry conditions for the $\Mink{D}$ solution—derived from spinor bilinears and G-structures—match those of the parent geometry, with a modified Bianchi identity. 

Using this framework, we construct gravitational duals to the twisted compactification of the SQCD-like theories. Starting from the backgrounds of \cite{Casero:2006pt, Casero:2007jj, Hoyos-Badajoz:2008znk}, which are singular, we find that the new Type II backgrounds constructed using \cite{Macpherson:2024qfi} are regular {\it only if} $N_f = 2N_c$.

This is a key result: first, the twisted compactification on S$^1$ requires a non-anomalous R-symmetry \cite{Cassani:2021fyv}, which in this class of QFTs occurs exclusively when $N_f = 2N_c$ \cite{Casero:2006pt, Casero:2007jj, Hoyos-Badajoz:2008znk}. 
On the other hand, this is precisely the only case in which  the generating technique of \cite{Macpherson:2024qfi} yields a non-singular and reliable Type II background, correctly describing the twisted circle compactification.

\subsection{Organisation of this paper}
\begin{itemize}
    \item{ In Section \ref{sec:G-structurereview}, we  review the G-structure conditions for Type IIB backgrounds with $F_{3}$ flux, focusing on the case of D5-branes on S$^{2}$ with smeared flavour branes. Then, we give a brief description on the SQCD-like theories dual to these backgrounds, and comment on two solutions that have an exact U(1) R-symmetry. Aspects of the four-dimensional SQCD-like theory are succinctly discussed}
    \item{Section \ref{sec:Compactification} begins with a brief mention to QFT aspects of the compactified field theory. After that, the background dual to the field theory on Minkowski$_3\times \text{S}^1$ is constructed. Serendipitously, we show that the non-anomalous condition for the U(1) R-symmetry, $\Nf=2\Nc$, appears when requiring regularity of the geometry at the end of the space. The generated background is clearly written. BPS equations and series expansions for the warp factors are quoted. Some exact solutions mentioned.}
    \item{Section \ref{sec:Observables} is dedicated to the study of different observables. The Chern-Simons level, Wilson loops and the holographic central charge  are studied. Some calibrated cycles are also discussed.}
\end{itemize}

\section{Review of G-structures for D5-branes on \texorpdfstring{S$^{2}$}{S2}}\label{sec:G-structurereview}
In this section we briefly review the formalities of supergravity backgrounds with a four dimensional Minkowski space-time and internal six manifold preserving minimal SUSY.
The solutions in IIB with a Mink$_4$ factor decompose as
    \begin{equation} \label{Mink4}
        ds^{2} = e^{2A} ds^{2}(\Mink{4})+ds^{2}(\text{M}_6), \quad 
        F_{-}= f_{-} + e^{4A}\text{vol}_{4} \wedge \star_{6} \lambda (f_{-}),
    \end{equation}
where $(e^{A},f_{-})$, the NS 3-form $H$ and dilaton $\Phi$ have support on M$_6$ only and $\lambda(X_n)=(-1)^{[\frac{n}{2}]}$.  

When such a background preserves supersymmetry this can be phrased in terms of differential constraints on spinor bi-linears defined on M$_6$, namely
    \begin{subequations}\label{eq:mink4}
    \begin{align}
        &(d-H\wedge)(e^{3A-\Phi}\Psi_-)=0,\label{eq:mink4bps1}\\
        &(d-H\wedge)(e^{2A-\Phi}\text{Re}\Psi_+)=0,\label{eq:mink4bps2}\\
        &(d-H\wedge)(e^{4A-\Phi}\text{Im}\Psi_+)=\frac{1}{8}e^{4A}\star_6\lambda (f_{-}),\label{eq:mink4bps3}
    \end{align}
    \end{subequations}
		which are the conditions in string frame. When these  hold it can be proven \cite{Martucci:2005ht,Koerber:2007hd} that to have a solution of type II supergravity one need only impose that the Bianchi identities of the NS 3-form and magnetic portion of the RR poly-form hold, \textit{i.e}
\beq
dH=0,~~~~ (d-H\wedge )f_{-}=\Xi,\label{eq:seedbi}
\eeq
where $\Xi$ is a polyform parameterising potential source terms. If sources are present, they must be ``calibrated''. For sources that are extended in Mink$_4$ and wrap some even dimensional $k$-cycle, the following equality should hold
\beq
\sqrt{\det ( g_{MN}+B_{MN})} d\xi^k= \pm 8\text{Im}\Psi_k, \label{eq:mink4cal}
\eeq
where the pull back onto the $k$-cycle  should be understood and the $\pm$ distinguishes between sources and anti-sources. This is equivalent to solving the $\kappa$ symmetry constraints for such a source.

When M$_6$ supports an  SU(3)-structure\footnote{It is also possible for M$_6$ to support an SU(2)-structure but we will not consider this here.} the bi-linears decompose as
\beq
\Psi_+= \frac{1}{8} e^{i\alpha}e^{-i J},~~~~\Psi_-= \frac{1}{8} \Omega\label{eq:SU3bi}
\eeq
where the function $\alpha$ is a possibly point dependent phase  and $(J,\Omega)$ are a real 2-form and holomorphic 3-form, which satisfy
    \begin{equation}
        J\wedge \Omega = 0, \quad 
        \frac{1}{3!}J\wedge J \wedge J = \frac{i}{2^{3}}   \Omega \wedge\bar{\Omega}= \text{vol}(\text{M}_{6}),
    \end{equation}
We focus on the $\alpha = 0$ case and with this choice, the system \eqref{eq:mink4} reduce to the following three purely-geometric equations
\begin{equation}\label{BPSeqs123}
         d\left( e^{2A-\Phi} \right) = 0, \quad
         d\left( J \wedge J \right) = 0, \quad
         d\left( e^{3A-\Phi} \Omega \right) = 0,
    \end{equation}
and the following definition of the 3-form flux
    \begin{equation}\label{BPSeqs4}
        f_{3} = - e^{-4A} \star_6 d \left( e^{4A-\Phi} J  \right).
 \end{equation}
The other fluxes are set to zero $f_{1} = f_{5} = H = 0$. For the case at hand we have
\beq
F_-=F_3=f_-=f_3.
\eeq
Let us now apply this information for the case of holographic duals to SQCD-like theories.
\subsection{Duals to \texorpdfstring{$\mathcal{N}=1$}{N=1} SQCD-like Theories}\label{sec:D5onS2}

We start with the following ansatz for the internal manifold
    \begin{equation} \label{eq:metricM6}
        ds^{2} (\text{M}_{6}) = e^{2A}\left( e^{2\tk} dr^{2} 
    + e^{2h} ds^{2}(\text{S}^2) + \frac{e^{2g}}{4}((\omega_1-{\cal A}_1)^2+ (\omega_2-{\cal A}_2)^2)+\frac{e^{2k}}{4}(\omega_3-{\cal A}_3)^2\right),
    \end{equation}
where
    \begin{equation}
        ds^2(\text{S}^2)= d\theta_1^2+\sin^2\theta_1 d\phi_1^2, \quad
        \calA_1=-a(r) \, d\theta_1, \quad 
        \calA_2 = a(r) \, \sin\theta_1 d\phi_1, \quad 
        \calA_3=-\cos\theta_1 d\phi_1,\label{oneform}
    \end{equation}
the 1-forms $\omega_i$ are SU(2) left invariant forms on the 3-sphere
\begin{equation}
\begin{aligned}
    &\omega_1=\cos\psi d\theta_2+\sin\theta_2\sin\psi d\phi_2, \\
    &\omega_2=-\sin\psi d\theta_2 +\sin\theta_2\cos\psi d\phi_2, \\ 
    &\omega_3=d\psi +\cos\theta_2 d\phi_2,
\end{aligned}
\end{equation}
which obey
\begin{equation}
    d\omega_i+\frac{1}{2}\epsilon_{ijk}\omega_j\wedge \omega_k = 0,
\end{equation}
and $(a,\tk,k,h,g,A)$ and the dilaton are functions of $r$ only. 
With the choice of vielbein
\begin{equation}
\begin{aligned}
    & e^r = e^{A+\tk}dr, \quad 
    e^{1} = e^{A+h} d\theta_2,\quad 
    e^{2} = e^{A+h} \sin\theta_1 d\phi_1,\\ 
    & e^{\hat{1},\hat{2}} = \frac{e^{A+g}}{2}(\omega_{1,2} -\calA_{1,2}),\quad 
    e^{\hat{3}} = \frac{e^{A+k}}{2}(\omega_{3}-\calA_{3})
\end{aligned}
\end{equation}
the SU(3)-structure forms are defined as 
\begin{align}
J&= e^{r}\wedge e^{\hat{3}}- e^{1}\wedge (\cos\beta e^{2}+\sin\beta e^{\hat{2}})-e^{\hat{1}}\wedge (\cos\beta e^{\hat{2}}-\sin\beta  e^{2}),\nn\\[2mm]
\Omega&= (e^r+i e^{\hat{3}})\wedge (\cos\beta e^{2}+\sin\beta e^{\hat{2}}+i e^{1})\wedge (\cos\beta e^{\hat 2}-\sin\beta e^{2}+i e^{\hat{1}}).\label{eqn:JandOmega}
\end{align}
where $0<\beta(r)<\pi/2$ is a rotation in the $e^2-e^{\hat{2}}$ plane.

The family of solutions of this system (provided $dF_{3}=0$) are dual to theories that flow to 4D $\mathcal{N}=1$ Super-Yang-Mills \cite{Maldacena:2000yy}. In \cite{Casero:2006pt,Casero:2007jj,Hoyos-Badajoz:2008znk,Gaillard:2010qg,Nunez:2010sf,Conde:2011aa, Conde:2011ab, Caceres:2011zn}, this interpretation is discussed at length and the addition of degrees of freedom transforming in the fundamental of the gauge group (quarks) in the form of additional brane sources is studied.

In fact, it is possible to consider the backreaction of $\Nf$ flavour D5-branes, extended along $(\Mink{4},r,\psi)$ and smeared on $(\theta_1,\phi_1,\theta_{2},\phi_{2})$. This configuration of the flavour branes preserves the supersymmetry of the system as they are $\kappa$-symmetric \cite{Nunez:2003cf}. See \cite{Nunez:2010sf} for a review of this procedure in different set-ups.

 The backreaction of the flavour branes does not modify the BPS equations\footnote{We note that although the differential form equations \eqref{BPSeqs123}-\eqref{BPSeqs4} do not change in their form, the ODEs obtained from them are different when including flavours.} \eqref{BPSeqs123} and \eqref{BPSeqs4}, but does modify the Bianchi identity to 
    \begin{equation}\label{SourcedBianchi}
        dF_{3} = \frac{\Nf}{4} \sin\theta_1\sin\theta_2d\theta_1 \wedge d\phi_1 \wedge d\theta_{2} \wedge d\phi_{2}. 
    \end{equation}

Solution of the system \eqref{BPSeqs123}-\eqref{BPSeqs4} which also satisfy the sourced Bianchi identity \eqref{SourcedBianchi}, solve the Type IIB equations of motion,  with sourced dilaton, and three form $F_3$--see \cite{Casero:2006pt}. 
These equations are modified to include the stress-energy tensor of the smeared branes, see the review \cite{Nunez:2010sf}. 

%
In summary, the models discussed above consist of a stack of $N_c$ D5 branes wrapped on a two-cycle inside the resolved conifold. To these, a stack of calibrated $N_f$ D5 branes is added. Since $N_f\sim N_c$ the backreaction must be considered. 

Having presented the basics of the supergravity backgrounds, we now discuss basic aspects of the dual field theories.
\subsubsection{Rudiments of the dual SQCD-like theory}

It was proposed in \cite{Casero:2006pt, Casero:2007jj, Hoyos-Badajoz:2008znk, Nunez:2010sf} that backgrounds like those analysed above are dual to ${\cal N}=1$ SQCD-like theories. 

In particular a perturbative Lagrangian was proposed.
In 4d ${\cal N}=1$ notation, the intervening fields are: a massless vector multiplet $W_\alpha$, a tower of massive vector fields $V_k$ and massive chiral multiplets $\Phi_k$. These fields transform in the adjoint of the gauge group SU$(N_c)$. The Lagrangian describing the dynamics of these fields is given in \cite{Andrews:2005cv, Andrews:2006aw}. These papers carefully work out the (twisted) compactification to four dimensions of a D5 brane wrapping the two cycle inside the resolved conifold.

On top of the dynamics above, the D5-sources 
extending on four dimensional Minkowski, the radial direction and the $\psi$-direction add new fields, which transform in the fundamental representation of SU$(N_c)$. In fact, we have massless quark superfields $Q,\tilde{Q}$. Other massive fields in the fundamental representation are expected to arise.

Of all the interactions between the massive adjoint matter and the massless quarks, we single out a superpotential term ${\cal W}\sim \tilde{Q}\Phi_k Q$. For details see \cite{Casero:2006pt, Casero:2007jj, Hoyos-Badajoz:2008znk, Nunez:2010sf}.

It is then natural to integrate out the massive matter. This procedure should be taken with a grain of salt: the mass of the KK-modes is degenerate with the gap-scale of the field theory. It was suggested in \cite{Casero:2006pt, Casero:2007jj, Hoyos-Badajoz:2008znk}
that the Lagrangian after this, schematically reads,
\begin{equation}
    \mathcal{L}_{\text{eff}}= \int d^4\theta~ \tilde{Q}^\dagger e^{V}\tilde{Q}  +Q^\dagger e^{-V}Q +\int d^2\theta ~W_\alpha W^\alpha+ W_{\text{eff}}.\label{efflag}
\end{equation}
The superpotential is
\begin{equation}
    W_{\text{eff}}\sim \frac{1}{\hat{\mu}}\left(\text{Tr} (\tilde{Q}Q)^2 -\frac{1}{N_c}(\text{Tr} \bar{Q} Q)^2 \right) +\hat{\lambda}(Q\tilde{Q})^3 +\tilde{z} (Q\tilde{Q})^4+....\label{superpotential}
\end{equation}
At high energies, the system is UV completed by the twisted compactification of the ${\cal N}=(1,1)$ Little String Theory with smeared D5-sources.


The SQFT in eqs.(\ref{efflag})-(\ref{superpotential}) has the global symmetries SU$(N_f)_V\times$U$(1)_B\times$U$(1)_\text{R}$. Only the vectorial part of the chiral symmetry SU$(N_f)_V\times$SU$(N_f)_A$ of SQCD survives. The baryonic symmetry of SQCD  U$(1)_B$ and  U$(1)$ R-symmetry  are present in our system. It is possible to assign charges to the different multiplets in the low energy description of (\ref{efflag}), see \cite{Casero:2006pt,Casero:2007jj, Hoyos-Badajoz:2008znk}. The charges of the chiral multiplets under the global symmetries are given in Table \ref{Table:Rcharges}.

\begin{table}
        \centering
        \begin{tabular}{c|cccc}
                  &  $\text{SU}(\Nf)_{V}$ & $\text{U}(1)_{B}$ & $\text{U}(1)_{\text{R}}$  \\ \hline     
            $Q$   &     $\Nf$      &    $1$     & $\frac{1}{2}$ \\
            $\tilde{Q}$ &     $\Nf$      &    $-1$    & $\frac{1}{2}$ 
        \end{tabular}
        \caption{Matter content and charges of the SQCD-like theory}
        \label{Table:Rcharges}
    \end{table}
    
The R-charge of the fermionic components is $\calR[\psi_{Q}]=\calR[Q]-1$ while the gaugino has the canonical R-charge $\calR[\lambda]=1$, is uncharged under U(1)$_{B}$ and a singlet of SU$(\Nf)_{V}$.

We are interested in working with SQFTs with non-anomalous R-symmetry. This is equivalent to the vanishing of the triangle diagram of three currents $<J_L(x)J_L(y)j_g(z)>$, consisting of two gauge currents $J_L$ and one global current $j_g$. This can also be phrased as the vanishing of the $\theta$-angle under fermionic chiral rotation ($\psi\to e^{i \epsilon \gamma_5} \psi$). This is
\begin{equation}
 \Delta\theta=\epsilon \sum_r n_r T(r)\calR[\psi].  
\end{equation}
Here $n_r$ is the number of particles in the $r$-representation of the gauge group, $T(r)$ is the weight of the representation (remind that $T[\text{adj}]=2N_c$ and $T[\text{fund}]=1$). 
For the theory in (\ref{efflag}), with $N_f$ quarks in the fundamental, $N_f$ antiquarks in the anti-fundamental and a gluon superfield in the adjoint, we find
\begin{equation}
 \Delta\theta= \epsilon\left( 2 N_c + 2\times  (-\frac{1}{2})\times N_f\right)=\epsilon (2N_c -N_f).\label{anomalyfree}   
\end{equation}
Then,  the SQFT has a non-anomalous R-symmetry only when 
\begin{equation}
\boxed{\colorbox{white!20}{$ N_f=2N_c. $}}\label{nf-2ncg}
\end{equation}
In  what follows, we discuss the dual geometric side of this version of SQCD. We write BPS equations and present solutions (which turn out to be singular). Further down the paper, we resolve the singularity by adding a gap on the QFT, capping of the geometry using the methods of \cite{Macpherson:2024qfi}. 
\subsection{Two exact solutions preserving a U(1) R-symmetry}\label{sec:Mink4U(1)}

We have argued that a non anomalous U(1) R-symmetry is required to compactify our field theory, however generically the background of \eqref{eq:metricM6} does not support such a symmetry, anomalous or otherwise. For $\partial_{\psi}$ to be a  U(1) isometry  it is necessary to impose that
\beq
a=0,~~~~\beta=0,
\eeq
the later being a consequence of the former, as it is the  $\psi$-direction that realises the R-symmetry. This means that the our internal space becomes a foliation of $\mathbb{T}^{(1,1)}$ over an interval of the form
\beq
e^{-2A}ds^2(\text{M}_6)=e^{2\tilde{k}}dr^2+  e^{2h}ds^2(\text{S}^2_1)+ \frac{e^{2g}}{4}ds^2(\text{S}^2_2)+ \frac{e^{2k}}{4}D\psi^2,\label{eq:M6def}
\eeq
where
\beq
ds^2(\text{S}^2_{i})= d\theta_i^2+ \sin^2\theta_i d\phi_i^2,~~~i=1,2,~~~~~ D\psi=d\psi+ \cos\theta_1 d\phi_1+ \cos\theta_2 d\phi_2,
\eeq
which preserves an SU(2)$\times$SU(2)$\times$U(1) isometry. The most general 3-form flux compatible with this isometry and the non zero terms that can be generated by \eqref{BPSeqs4} given our restricted ansatz is
\beq
F_3=f_3= \bigg(u_1 \text{vol}(\text{S}^2_1)+ u_2 \text{vol}(\text{S}^2_2)\bigg)\wedge D\psi,\label{eq:f3def1}
\eeq
where $u_{1,2}$ are functions of $r$, generically. Inserting this ansatz into \eqref{BPSeqs123}-\eqref{BPSeqs4} leads to the following BPS conditions that imply supersymmetry
\begin{subequations}
\begin{align}
&\partial_r(e^{2A-\Phi})=0,\label{eq:bpsequations1}\\[2mm]
&e^{2A-\Phi}\partial_r\left(e^{2h}-\frac{1}{4}e^{2g}\right)=2 e^{k-\tilde{k}}(u_1-u_2),\label{eq:bpsequations2}\\[2mm]
&\partial_{r}(e^{2(h+g-k)})=e^{\tilde{k}-3k}\left(4 e^{2(h+k)}+e^{2g}(e^{2k}-4 e^{2h})\right),\label{eq:bpsequations3}\\[2mm]
&\partial_r(e^{2A+h+g})=e^{2A+\tilde{k}+k-g-h}\left(e^{2g}+\frac{1}{4}e^{2h}\right),\label{eq:bpsequations4}\\[2mm]
&e^{2A-\Phi}\partial_r(e^{2h})=\frac{1}{2}e^{2A-\Phi+\tilde{k}+k}+2 e^{\tilde{k}-k}u_1.\label{eq:bpsequations5}
\end{align}
\end{subequations}
Of course these only imply a solution off shell. Consistency with the Bianchi identity \eqref{SourcedBianchi} demands that we fix $(u_1,u_2)$ to be constant and when this also holds the rest of the type IIB equations of motion follow. In particular we must fix
\beq
4u_1= N_c-N_f,~~~~4 u_2=-N_c\label{eq:udefs1}
\eeq
such that 
\beq
-\frac{1}{(2\pi)^2}\int F_3=N_c,
\eeq
with the integration cycle being the 3-sphere spanned by $(\theta_2,\phi_2,\psi)$ evaluated at constant $(\theta_1,\phi_1)$. 
Without loss of generality we solve \eqref{eq:bpsequations1} as
\beq
e^{2A}= e^{\Phi},\label{eq:eArule}
\eeq
where a possible integration constant can be absorbed into the other functions appearing in the metric and by a rescaling of the coordinates on Mink$_4$.
Solutions then take the form
\begin{align}
ds^2&= e^{\Phi}\bigg[ds^2(\text{Mink}_4)+ e^{2\tilde{k}}dr^2+  e^{2h}ds^2(\text{S}^2_1)+ \frac{e^{2g}}{4}ds^2(\text{S}^2_2)+ \frac{e^{2k}}{4}D\psi^2\bigg],\nn\\[2mm]
F_3&=\frac{1}{4}\bigg((N_f-N_c) \text{vol}(\text{S}^2_1)-N_c \text{vol}(\text{S}^2_2)\bigg)\wedge D\psi.
\end{align}
Solutions are governed by
the  equations
\begin{subequations}
\begin{align}
&\partial_r\left(e^{2h}-\frac{1}{4}e^{2g}\right)=\frac{1}{2}e^{k-\tilde{k}}(2N_c-N_f),\label{eq:so11}\\[2mm]
&\partial_{r}(e^{2(h+g-k)})=e^{\tilde{k}-3k}\left(4 e^{2(h+k)}+e^{2g}(e^{2k}-4 e^{2h})\right),\label{eq:sol2}\\[2mm]
&\partial_r(e^{\Phi+h+g})=e^{\Phi+\tilde{k}+k-g-h}\left(e^{2g}+\frac{1}{4}e^{2h}\right),\label{eq:sol3}\\[2mm]
&\partial_r(e^{2h})=\frac{1}{2}e^{\tilde{k}+k}+\frac{1}{2} e^{\tilde{k}-k}(N_c-N_f).\label{eq:sol4}
\end{align}
\end{subequations}
Despite $\partial_{\psi}$ being a Killing vector of the entire background of this section, it can be argued geometrically that it is anomalous unless one fixes $\Nf = 2\Nc$. Indeed, a euclidean D1-brane probe wrapping a two cycle (representing an instanton in the QFT), was found to have a theta-term proportional to $(N_f-2N_c)$, see \cite{Casero:2006pt, Casero:2007jj}. This argument is the holographic version of the anomaly cancellation leading to \eqref{nf-2ncg}. In Section \ref{sectionIRregular} we discuss a modification of these backgrounds. We show that to achieve a smooth modified geometry imposes $\Nf = 2\Nc$.

In the case of $\Nf=2\Nc$ there exist two exact solutions which will be relevant to the rest of this work, that yield respectively a linear and asymptotically constant dilaton as $r\to \infty$. The asymptotically linear dilaton solution was found in \cite{Casero:2006pt} and solves \eqref{eq:so11}-\eqref{eq:sol4}. It reads,
\beq
e^{2\tilde{k}}=e^{2k}=N_c,~~~~e^{2h}=\frac{N_c}{\xi}~~~~e^{2g}=\frac{4N_c}{4-\xi},~~~~e^{\Phi-\Phi_0}= e^{r},\label{eq:lindil}
\eeq
for which $r\in \mathbb{R}$ and $(\xi,\Phi_0)$ are constants with the former constrained as $0<\xi<4$. There is a singularity at $r=-\infty$ where the dilaton is vanishing. The asymptotically constant dilaton solution was found in \cite{Hoyos-Badajoz:2008znk} and is defined as
\begin{align}
e^{2\tilde{k}}&=e^{2k}=\frac{2}{3}\Delta_1,~~~~e^{2h}=\frac{1}{4}\Delta_1~~~~e^{2g}=\Delta_2~~~~e^{\Phi-\Phi_0}= \frac{6^{\frac{1}{4}}e^{r}}{\sqrt{\Delta_1}\Delta_2^{\frac{1}{4}}},\nn\\[2mm]
\Delta_1&=\frac{3N_c}{2}+ c_+ e^{\frac{4}{3}r},~~~~\Delta_2=3N_c+ c_+ e^{\frac{4}{3}r},\label{eq:constdil}
\end{align}
for $(c_+,\Phi_0)$ integration constants and where again $r \in \mathbb{R}$. This solution approaches Mink$_4\times {\cal C}_6$ as $r\to \infty$ for ${\cal C}_6$ a CY$_3$ cone (at infinity) whose base is the Sasaki-Einstein metric on $\mathbb{T}^{(1,1)}$. As such the solution is regular about $r=\infty$, however again there is a singularity at $r=-\infty$ where $e^{\Phi}$ is vanishing.

In the next section we  construct solutions that compactify Mink$_4$ to Mink$_3\times \text{S}^{1}$ that share the same  UV asymptotic form as the above two solutions, but enjoy regular behaviour in the IR.

\section{Compactification from \texorpdfstring{Mink$_4$ to Mink$_3$}{Mink4 to Mink3}.}\label{sec:Compactification}
Here, we study the twisted circle compactification of the theories described above. We start by discussing the field theory side, and then move to the construction of its holografic dual, for which we follow \cite{Macpherson:2024qfi}. As mentioned before, in order to properly describe the circle compactification, these geometries need a smoothly shrinking S$^{1}$, which is fibered over the U(1) R-symmetry direction. 

The condition $N_f=2N_c$ appears when imposing regularity of the newly constructed background. This closes a nice circle of ideas: the non-anomalous character
of the R-symmetry (needed to perform the twist) imposes the relation $N_f=2N_c$, in agreement with the regularity of the backgrounds dual to the compactified SQCD.

We start the section with a discussion of the compactified field theory and then elaborate on the string dual.
\subsection{Field Theory Perspective}

Following \cite{Cassani:2021fyv}, it is only possible to perform a further SUSY-preserving, twisted-compactification on a Minkowski-circle in the case in which the R-symmetry is preserved. This is (\ref{nf-2ncg}),
\begin{eqnarray}
\text{Mink}_4 \underset{\text{SUSY}}{\longrightarrow} \text{Mink}_3\times \text{S}^1  \longleftrightarrow N_f=2N_c .\nonumber 
\end{eqnarray}
Let us study the effects of the twisted-circle compactification. One of the Minkowski directions is compactified on a circle of radius $R_0$ parametrized by $\varphi$. Denoting by $\Psi$ a generic field, we impose boundary conditions as 
    \begin{equation}
    \Psi(x,\varphi) = \sum_{n\in \mathbb{Z}} e^{\frac{ i \varphi}{2\pi R_0}(n+\mathfrak{s})} \Psi^{(n)}(x),\label{compact-1}
    \end{equation}
where $\mathfrak{s} =0$ for bosons and $\mathfrak{s} =1/2$ for fermions. 

To couple the theory to the background gauge field for the U(1) R-symmetry, we modify the gauge covariant derivative as (see \cite{Kumar:2024pcz} for a more detailed discussion on a different theory)
    \begin{equation}
        D_{\mu}\Psi  \rightarrow \calD_{\mu}\Psi = D_{\mu}\Psi - i \calR[\Psi] \hat{\calA}_{\mu}\Psi,\qquad \hat{\calA}_\mu = Q \delta^{\varphi}_{\mu}\,,
    \end{equation}
where $\calR[\Psi]$ is the R-charge of the field. This coupling shifts the effective 3D mass of the KK modes on the circle as
    \begin{equation}
        m_{\Psi^{(n)}} = \frac{1}{R_0}\left| n + \mathfrak{s}  \right| \rightarrow \left| \frac{1}{R_0}\left(n + \mathfrak{s} \right) - \calR[\Psi] Q  \right|.
    \end{equation}
Notice that $Q=0$ breaks supersymmetry since there would be no massless zero mode for the fermionic gauge superpartner. Since we want to retain the $\mathcal{N}=1$ massless vector multiplet it is convenient to set 
    \begin{equation}
        Q = \frac{\mathfrak{s}}{R_0}\frac{1}{\calR[\lambda]} = \frac{1}{2R_0} \,.
    \end{equation}
With this choice, we recover a vector-like spectrum for the circle KK-modes of gaugino, but the mass spectrum of the chiral multiplet gets modified as
    \begin{equation}
        m_{Q} = \frac{1}{R_0}\left|n-\frac{1}{4}\right|, \quad
        m_{\psi_{Q}} = \frac{1}{R_0}\left|n-\frac{1}{4}\right|.\label{compact-4}
    \end{equation}
So even if the mass spectrum is not vector-like, we manage to restore Bose-Fermi degeneracy. Finally, at scales lower than the radius of the circle, we can integrate out the KK-modes. When doing so, there is a Chern-Simons level generated due to the integration of the fermions \cite{Cassani:2021fyv}
    \begin{equation}\label{eq:3DCSQFT}
        k = \Nf \sum_{n \in \mathbb{Z}} \text{sign}\left( n - \frac{1}{4} \right) = \Nc.
    \end{equation}
Below we study the holographic dual to the twisted compactification of this SQCD-like theory with dynamics in \eqref{efflag}.  
%
We are  interested in finding the holographic  dual to the procedure of compactifying the field theory on a circle with a twist of the R-symmetry, as discussed in eqs.(\ref{compact-1})-(\ref{compact-4}). 

\subsection{Supergravity Duals}

This kind of dual supergravity solution has been the subject of intense study recently, see for example \cite{Anabalon:2021tua,Bobev:2020pjk,Anabalon:2022aig,Nunez:2023nnl,Nunez:2023xgl, Fatemiabhari:2024aua,Chatzis:2024top,Chatzis:2024kdu,Anabalon:2024che,Castellani:2024pmx,Fatemiabhari:2024lct, Kumar:2024pcz, Castellani:2024ial}. To achieve the compactification we follow the general procedure of \cite{Macpherson:2024qfi} and  modify the ansatz of \eqref{Mink4} as
    \begin{equation}
        ds^2(\text{Mink}_{4}) \rightarrow   ds^2(\text{Mink}_{3})+ D\varphi^2, \quad D\varphi=d\varphi+\calA, \quad 
        \calF = d\calA,
    \end{equation}
where $\calF$ has support on M$_{6}$ and $\partial_{\varphi}$ is an isometry of the background. The result can be viewed as a Mink$_3$ background whose internal space is a U(1) fibration over M$_6$, \textit{i.e.}
\beq
ds^{2}(\text{M}_{7}) = e^{2A}D\varphi^2 + ds^{2}(\text{M}_{6}).
\eeq
In \cite{Macpherson:2024qfi} it was proven that if M$_6$ supports an SU(3)-structure in type IIB as described in Section \ref{sec:G-structurereview}, then the only additional condition that one needs to impose for such a circle compactification to preserve supersymmetry is that ${\cal F}$ is a primitive (1,1)-form, which is to say it obeys
    \begin{equation}
        {\cal F}\wedge J\wedge J={\cal F}\wedge \Omega=0.\label{ConditionsCalF}
    \end{equation}
This leads to a class of solutions that support an SU(3) structure of the form
    \begin{align}
        ds^2&=e^{2A}\left( ds^2(\text{Mink}_3)+D\varphi^2 \right)+ds^2(\text{M}_6),\nn \\[2mm]
        F_{-}&=\left(1+\star\lambda\right)\left(f_{-}+ \frac{8}{c_0}e^{2A-\Phi} D\varphi\wedge \calF\wedge\text{Re}\Phi_{+}\right),\label{TCCBackground}
    \end{align}
		with the NS flux $H$ of the same form as it was for Mink$_4$. Thus given a Mink$_4$ background that preserves supersymmetry we can easily construct a circle compactified background that also preserves supersymmetry.
One should appreciate however that this is only a solution generating technique off shell, in particular the Bianchi identities  of the RR fluxes in the parent Mink$_4$ background \eqref{eq:seedbi} should not be imposed, instead this is modified as
\begin{equation}
(d-H\wedge)f_{-}+ 8{\cal F}\wedge {\cal F}\wedge(e^{2A-\Phi}\text{Re}\Psi_{+})=\Xi,\label{eq:bianchimod}
\end{equation}
where again $\Xi$ parametrises possible source terms.

Such solutions preserve supersymmetry in terms of bi-spinors on M$_7$ defined in terms of \eqref{eq:SU3bi} as
\beq
\Psi^{(7)}_+=-\text{Re}(\Psi_++e^{A}\Psi_-\wedge D\varphi),~~~~\Psi^{(7)}_-=\text{Im}(\Psi_--e^{A}\Psi_+\wedge D\varphi),
\eeq 
which generically realise an SU(3)-structure but which is enhanced to G$_2$-structure when $\alpha=0$, as is the case for the solutions we will consider. In this case the G-structure 3-form is defined as
\beq
\Phi_3= J\wedge D\varphi+\text{Im}\Omega.
\eeq
As with the case of Mink$_4$, for supersymmetry and the Bianchi identities to imply the remaining supergravity equations of motion it is necessary  when sources are present, that they are calibrated. For sources  extended in $(\text{Mink}_3)$ and wrapping some odd $k+1$-cycle this amounts to imposing
\beq\label{eq:3-cyclesusy}
\sqrt{\det ( g_{MN}+B_{MN})} d\xi^{k+1}= \mp 8\text{Im}\Psi^{(7)}_{k+1},
\eeq
where the pull back onto the $k+1$ cycle is understood. Note that for sources that wrap $\varphi$ this is implied by \eqref{eq:mink4cal}.

\subsection{Necessary conditions for solutions}
We apply the techniques described above to obtain the dual of the twisted circle compactification of SQCD like theories. As in Section \ref{sec:Mink4U(1)} we impose that $\partial_{\psi}$ is an isometry so that M$_6$ takes the form of \eqref{eq:M6def}. Supersymmetry still demands that we impose \eqref{eq:bpsequations1} such that M$_6$ does indeed support an SU(3)-structure, and we can still fix $e^{2A}$ as in \eqref{eq:eArule}. We make the following ansatz for the connection
    \begin{equation}
        \calA = p(r) D\psi,~~~\Rightarrow~~~~{\cal F}= p'dr\wedge D\psi-p\left(\text{vol}(\text{S}^2_1)+\text{vol}(\text{S}^2_2)\right)
    \end{equation}
		such that the manifold M$_7$, which is a U(1) fibration, preserves the isometries of the base manifold M$_6$. Imposing that ${\cal F}$ is primitive as in \eqref{ConditionsCalF} then amounts to requiring that
		\beq
		\partial_r\log p=-\frac{1}{2}\left(\frac{4}{e^{2g}}+ \frac{1}{e^{2h}}\right)e^{\tilde{k}+k}.\label{eq:bps2}
		\eeq
All that remains is to impose the Bianchi identity \eqref{eq:bianchimod} for
\beq
\Xi=\frac{\Nf}{4} \text{vol}(\text{S}^2_1)\wedge \text{vol}(\text{S}^2_2).
\eeq
We still have that $f_-=f_3$ which takes the form of \eqref{eq:f3def1}. We no longer impose \eqref{eq:udefs1}, but rather \eqref{eq:bianchimod}. This fixes 
\beq
4u_1=N_c-N_f+4 p^2,~~~4u_2=-N_c+ p^2.
\eeq	
Upon inserting these definitions into \eqref{eq:bpsequations1}-\eqref{eq:bpsequations5} it is possible to show that \eqref{eq:bps2} can be integrated as
\beq
p = p_{0} e^{-2\Phi-2g-2h}.  
\eeq
In this way, we arrive at a class of solutions of the form
\begin{align}
ds^2&= e^{\Phi}\bigg[ds^2(\text{Mink}_3)+D\varphi^2+ e^{2\tilde{k}}dr^2+  e^{2h}ds^2(\text{S}^2_1)+ \frac{e^{2g}}{4}ds^2(\text{S}^2_2)+ \frac{e^{2k}}{4}D\psi^2\bigg],\nn\\[2mm]
F_3&=\bigg((N_c-N_f+4 p^2) \text{vol}(\text{S}^2_1)+(p^2-N_c)\text{vol}(\text{S}^2_2)\bigg)\wedge D\psi+ D\varphi\wedge {\cal F},\nn\\[2mm]
D\varphi&= d\varphi+p D\psi,~~~~~D\psi=d\psi+ \cos\theta_1 d\phi_1+\cos\theta_2 d\phi_2,
\end{align}
that are defined by the following system of ODEs
\begin{subequations}
\begin{align}
&\partial_r\left(e^{2h}-\frac{1}{4}e^{2g}\right)=\frac{1}{2} e^{k-\tilde{k}}(2N_c-N_f),\label{eq:compsol1}\\[2mm]
&\partial_{r}(e^{2(h+g-k)})=e^{\tilde{k}-3k}\left(4 e^{2(h+k)}+e^{2g}(e^{2k}-4 e^{2h})\right),\label{eq:compsol2}\\[2mm]
&\partial_r(e^{\Phi+h+g})=e^{\Phi+\tilde{k}+k-g-h}\left(e^{2g}+\frac{1}{4}e^{2h}\right),\label{eq:compsol3}\\[2mm]
&\partial_r(e^{2h})=\frac{1}{2}e^{\tilde{k}+k}+\frac{1}{2} e^{\tilde{k}-k}(N_c-N_f+ p^2).\label{eq:compsol4}
\end{align}
\end{subequations}
Any solution to these equations defines a solution in type IIB supergravity that preserves as many supersymmetries as the Mink$_4$ parent, which in this case is  4 real supercharges, or ${\cal N}=2$ supersymmetry from the Mink$_3$ perspective. Note that we can still define a charge of colour D5 branes as
\beq
-\frac{1}{(2\pi)^2}\int F_3=N_c,
\eeq
integrated over $(\theta_2,\phi_2,\psi)$ at constant $(\varphi,\theta_1,\phi_1)$. Let us discuss solutions to (\ref{eq:compsol1})-(\ref{eq:compsol4}).

\subsection{An IR expansion and regularity}\label{sectionIRregular}
We want to find a solution such that the U(1) direction $\varphi$ shrinks to zero size smoothly at some point $r=r_{0}$. In order to see how this is possible, we rewrite the line element as
    \begin{align} \label{eq:FinalAnsatz}
        ds^{2} &= e^{\Phi}\bigg[ ds^{2}(\Mink{3}) + \frac{e^{2k}}{e^{2k} + 4p^{2}}d\varphi^{2} + e^{2\tk} dr^{2} + e^{2h} ds^{2}(\text{S}^{2}_1)+ \frac{e^{2g}}{4}ds^2(\text{S}^2_2) + \left( p^{2} + \frac{e^{2k}}{4} \right){\cal D}\psi^2\bigg],\nn\\[2mm]
{\cal D}\psi&=D\psi + {\cal B},~~~~{\cal B}=\left(\frac{p}{p^{2}+\frac{e^{2k}}{4}} - {\cal B}_{0}\right) d\varphi .
    \end{align}
We also performed the coordinate transformation
    \begin{equation}
       \psi \rightarrow \psi - {\cal B}_{0}\varphi\label{eq:shift}
    \end{equation}
so that at $r=r_{0}$ we can tune ${\cal B}_{0}$ such that ${\cal B}$ is zero. This  requires
    \begin{equation}
         \frac{p}{p^{2}+\frac{e^{2k}}{4}} \bigg|_{r=r_{0}} - {\cal B}_{0} = 0.
    \end{equation}
If we can arrange for the $(r,\varphi)$ directions of the above metric to tend to the origin of  $\mathbb{R}^2$ in polar coordinates as $r\to r_0$ with the remaining warp factors constant then the space terminates regularly at this point. This gives the behaviour of a supersymmetric solitonic background, which are proposed to be duals to twisted circle compactification of field theories. We choose to employ a coordinate transformation in $r$ to fix
\beq
e^{\tilde{k}}= N_c e^{-k}
\eeq
such that the sub-manifold in question takes the form
\beq
ds^2(\text{M}_2)=e^{\Phi}\left(N_c^2e^{-2k} dr^{2}+\frac{e^{2k}}{e^{2k} + 4p^{2}}d\varphi^{2} \right).
\eeq
We can also use the invariance of \eqref{eq:compsol1}-\eqref{eq:compsol4} under $r\to r-r_0$ to assume that $r_0=0$. To realise a regular zero at $r=0$ we must assume  that $(e^{\Phi},e^{2g},e^{2h})$ are constant at leading order, which implies the same for $(p,{\cal B}_{\varphi})$, and that 
\beq
e^{k}= \sqrt{N_c}k_0 \sqrt{r}+...
\eeq
When this is true we have, 
\beq
ds^2(\text{M}_2)\sim \frac{dr^2}{4r}+  r \frac{1}{R_0^2}d\varphi^2,~~~~~R_0=\frac{4p(0)}{k_0^2}
\eeq
which indeed vanishes as the origin of $\mathbb{R}^2$ provided that $\varphi$ has period\footnote{One could of course be a bit more general and allow for a $\mathbb{Z}_k$ orbifold singularity, but we will not consider this possibility here.} $2\pi R_0$. Having established the behaviour we wish to have we must now examine its compatibility with the ODEs we need to solve. A regular zero means that the RHS of  \eqref{eq:compsol1} generically blows up as $\frac{1}{2}(N_c-2N_f) N_c^2k_0^2/r$ as a consequence we must have to leading order that
\beq
\frac{1}{4}e^{2g}-e^{2h}=-  \frac{1}{2}(2N_c-N_f) N_c^2k_0^2\log r,
\eeq
clearly this is incompatible with our assumption that $(e^{2h},e^{2g})$ are constant at this loci unless we tune $N_f=2N_c$. Likewise at this loci the RHS of  \eqref{eq:compsol4} blows up as $\frac{1}{r}(N_c-N_f+ p(0)^2)$ in a way that cannot be resolved with $e^{2h}$ being constant to leading order unless
\beq
(N_c-N_f+ p(0)^2)=0.
\eeq
This cannot be resolved with the previous condition unless $p \neq 0$. We thus find that a regular zero in M$_2$ demands that we fix
\beq
\boxed{\colorbox{white!20}{$ N_f=2N_c$}}
\eeq
and that we perform our compactification in the presence of a non trivial connection.

Let us highlight this result. From the dual field theory point of view, in order to perform the twisted circle compactification we need a non-anomalous R-symmetry, which in this case is achieved precisely when $\Nf=2\Nc$ and in the absence of spontaneous breaking to a discrete group. See the discussion around equation (\ref{anomalyfree}). It is notable that the supersymmetry conditions, Bianchi identity and requiring a smoothly shrinking field theory S$^{1}$ impose this constraint on the solution, which shows that even in classical gravity, where the anomaly is captured by a probe, smoothness of the background  requires the anomaly free condition.

Given that we fix $N_f=2N_c$ we can integrate \eqref{eq:compsol1} as
\beq
 e^{2g} = 4( N_c g_0 +  e^{2h})\label{eq:egdef}
\eeq
We now proceed to solve \eqref{eq:compsol2}-\eqref{eq:compsol4} in terms of the following series expansion  about $r=0$
    \begin{equation}
        e^{\Phi} = e^{\Phi_0}\left(1+\sum^{\infty}_{i=1} a_{i}\, r^{i}\right), \quad
        e^{h} = \sqrt{N_c}\left(\frac{1}{\sqrt{\xi}}+\sum^{\infty}_{i=1} h_{i}\, r^{i}\right), \quad
        e^{k} = \sqrt{N_c}\sqrt{r}\left(k_0+ \sum^{\infty}_{i=1} k_{i}\, r^{i}\right), \label{eq:IRseries}
    \end{equation}
where we assume $(\xi,k_0)$ are strictly non zero. We then find that we can recursively solve \eqref{eq:compsol2}-\eqref{eq:compsol4} order by order in $r$, we find that the leading order requires that we tune
\beq
k_0= 2,~~~~~p_0= 2 e^{2\Phi_0} \frac{1}{\xi} \left(g_0+ \frac{1}{\xi}\right)N_c^{\frac{5}{2}}
\eeq
while the next to leading order that
\begin{align}
a_1&= \frac{\left(g_0+\frac{2}{\xi}\right)^2}{16 \xi^4\left(g_0+ \frac{1}{\xi}\right)^2},~~~~h_1=\frac{\frac{4}{\xi^2}-\frac{2}{\xi}+g_0\left(\frac{4}{\xi}-1\right)}{16 \frac{1}{\xi^{\frac{3}{2}}}\left(g_0+\frac{1}{\xi}\right)},\nn\\[2mm]
k_1&= \xi^2\frac{\left(g_0+ \frac{2}{\xi}\right)\left(g_0+2(1+2 g_0)\frac{1}{\xi}+ \frac{4}{\xi^2}\right)}{16 \left(g_0+ \frac{1}{\xi}\right)^2},
\end{align}
the series persists in this fashion but we will not quote further terms. One interesting feature however is if we fix $h_1=0$ then the series for $e^{h}$ truncates at leading order - in this limit it is actually possible to solve the system of ODEs exactly leading to a solution we explore in section \ref{sec:exactsol}.

Given that $p(0)=\frac{\sqrt{N_c}}{2}$ we now find  that
    \begin{equation}
        {\cal B}_{0} = \frac{1}{R_0},~~~~~R_0=\frac{\sqrt{N_c}}{2},
    \end{equation}
where $2\pi R_0$ is the period of $\varphi$.  The 3-form flux may now be written as
\beq
F_3= -\frac{N_c}{4}(\text{vol}(\text{S}^2_1)+\text{vol}(\text{S}^2_2))\wedge D\psi-d\bigg((p-R_0)d\varphi\wedge D\psi\bigg)\label{eq:3-formaftershift}
\eeq
Given that $\partial_{\psi}$ is an R-symmetry direction it is reasonable to ask at this point whether the shift in \eqref{eq:shift} breaks supersymmetry globally. We can answer this question in terms of the holomorphic 3-form $\Omega$. Given its form in \eqref{eqn:JandOmega} it should be clear that for $\beta=0$ this contains the phase $e^{i\psi}$ clearly this is mapped into itself under $\psi\to \psi+ 4\pi$ as it should be. After the shift we have that
\beq
e^{i\psi}\to  e^{i(\psi- \frac{1}{R_0}\varphi)}.
\eeq
Things work as before for the new $\psi$ coordinate but  now we also need the shifted phase to be mapped to itself as  the entire $\varphi$ circle is traversed, \textit{i.e.} when $\varphi\to\varphi+ 2\pi R_0$ which the $\frac{1}{R_0}$ factor ensures. Thus there is no global issue with the shift in \eqref{eq:shift}, however the spinor the background supports is  anti periodic in $\varphi$ rather than periodic as it is in $\psi$.

In the next section we  show that it is also possible to derive a UV series expansion that gives rise to an asymptotically constant dilaton and further, we numerically interpolate between it and the IR expansion of this section.

\subsection{Asymptotically constant dilaton and numerical interpolation}
Having found a series expansion that gives rise to a regular IR we would like to connect it to the UV behaviours of Section \ref{sec:Mink4U(1)}. This is a two step process. First we attempt to find a series expansion realising either an asymptotically linear or asymptotically constant dilaton. Next one confirms that it is indeed possible to numerically interpolate between the IR of the previous section and the UV series. We  focus on asymptotically constant dilaton in this section, as we find an exact solution with a linear dilaton in the next section that we have not been able to further generalise in terms of a UV series.

Again fixing $(e^{\tilde{k}},e^{2g})$ as in the previous section we find that  eqs.\eqref{eq:compsol2}-\eqref{eq:compsol4} are solved by the following asymptotic expansions
\begin{align}\label{eq:ExpansionUVConstantDilaton}
e^{\Phi}&=e^{\Phi_{\infty}}\bigg(1-\frac{3}{4 r}- \frac{\left(b_+-\frac{9}{16}\log r\right)}{ r^2}+O\left(\frac{1}{r}\right)^3\bigg),\nn\\[2mm]
e^{h}&=\sqrt{\frac{N_c r}{2}}\bigg(1+\frac{\frac{1}{96}(9+64 b_+-48 g_0)-\frac{3}{8}\log r}{r}+O\left(\frac{1}{r}\right)^2\bigg),\nn\\[2mm]
e^{k}&=2\sqrt{\frac{N_c r}{3}}\left(1-\frac{\frac{1}{96}(9-64b_+)+\frac{3}{8}\log r}{r}+O\left(\frac{1}{r}\right)^2\right),
\end{align}
where $(\Phi_{\infty},b_+)$ are UV integration constants.
The series appears to persist indefinitely in this  fashion with an additional power of $\log r$ appearing at each higher order inverse power of $r$ in the expansion. Notice that $p_0$ does not appear in the expansions to the order we have written explicitly. This is because to leading order
\beq
p= p_0\frac{e^{-2\Phi_0}}{N_c^2 r^2},
\eeq
so its effect is heavily suppressed in the UV and it in fact first enters at $O\left(\frac{1}{r}\right)^5$ in the expansion of $e^{\Phi}$. There is no barrier in the UV to tuning $p_0$ as is required to have a regular IR.

It is not hard to confirm that about infinity the metric tends to 
\beq
e^{-\Phi_{\infty}}ds^2=  ds^2(\text{Mink}_3)+d\varphi^2+ 3N_c\bigg[\frac{dr^2}{4r}+ r\bigg(\frac{1}{6}ds^2(\text{S}^2_1)+\frac{1}{6}ds^2(\text{S}^2_2)+ \frac{1}{9}(D\psi-\frac{1}{R_0}d\varphi)^2\bigg)\bigg],
\eeq
which is locally a direct product on Mink$_4$ and a CY3 cone over $\mathbb{T}^{(1,1)}$, reproducing the UV behaviour of the solution of \eqref{eq:constdil}.

\begin{figure}[h]
\centering
\includegraphics[scale=0.5]{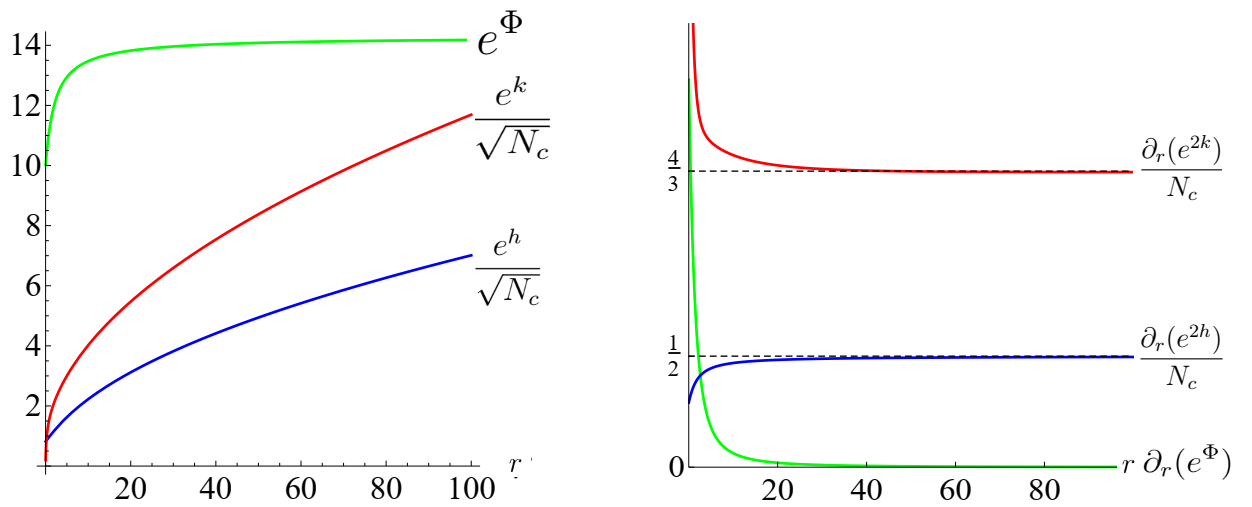}
\caption{A numerical plot of an asymptotically constant dilaton solution. The left plot shows the global properties of $(e^{\Phi},e^{h},e^{k})$, while the right plot confirms that \eqref{eq:limits} is indeed realised. In this example we have tuned the IR integration constants as $(e^{\Phi_0}=10,~\xi=1.5,~g_0=5)$.  }\label{fig:numerics}
\end{figure}

It remains to establish that the regular IR expansion of the previous section and the UV in this section are two limits of a common global solution, at least for some tunings of their respective integration constants. We can achieve this by numerically solving \eqref{eq:compsol2}-\eqref{eq:compsol4} with $N_f=2N_c$ taking the series expansion in \eqref{eq:IRseries} as the IR boundary conditions for $(e^{\Phi},e^{h},e^{k})$ then numerically interpolating towards the UV. We conclude that the IR and UV series can be interpolated between if the numerics reproduces the UV asymptotic expansion, this happens if the following expressions converge to
\beq
\partial_{r}e^{\Phi}\to 0,~~~~ \frac{\partial_{r}(e^{2h})}{N_c}\to \frac{1}{2},~~~~\frac{\partial_{r}(e^{2k})}{N_c}\to\frac{4}{3},\label{eq:limits}
\eeq 
we are able to confirm that this is indeed the case for certain tunings of $(\Phi_0,g_0,\xi)$, an example of this matching and the global properties of $(e^{\Phi},e^{h},e^{k})$ are given in Figure \ref{fig:numerics}. A more detailed study suggests that global solutions of this type exist for generic $\Phi_0$ and $(\xi,g_0)$, satisfying the inequalities
\beq
\xi>0,~~~~\xi g_0+1>0,~~~~4-2\xi+g_0(4-\xi)\xi>0
\eeq
the first two of which ensure positive real metric factors - it is really the last one that is relevant. Note that it is $4-2\xi+g_0(4-\xi)\xi=0$ that truncates the IR series for $e^{h}$  to leading order.

In the next section we present an exact solution with  globally linear dilaton which also connects to our regular IR expansion.

\subsection{An exact solution with linear dilaton}\label{sec:exactsol}
In this section we derive an exact solution with linear dilaton and no IR singularity.
We have already noted that fixing
\beq
g_0=\frac{2(2-\xi)}{(\xi-4)\xi}
\eeq
causes the IR expansion for $e^{h}$ to terminate at leading order where is constant. Thus we proceed by fixing $g_0$ as above and simply fixing
\beq
e^{h}=\sqrt{\frac{N_c}{\xi}},~~~p_0=\frac{2 e^{2\Phi_0}N_c^{\frac{5}{2}}}{(4-\xi)\xi}
\eeq
in \eqref{eq:compsol2}-\eqref{eq:compsol4} which become rather trivial and can be integrated as
\beq
e^{k}=\sqrt{N_c}\sqrt{f},~~~f=1- e^{4(\Phi_0-\Phi)},~~~~e^{\Phi-\Phi_0}=e^r,
\eeq
yielding an exact solution. The metric becomes 
    \begin{equation}\label{eq:ExactSolutionMetric}
    \begin{aligned}
        ds^{2} &= e^{\Phi}\bigg[ ds^{2}(\Mink{3}) + fd\varphi^{2} + N_{c}\bigg(\frac{dr^{2}}{f} + \frac{1}{\xi} ds^{2}(\text{S}^{2}_1)+\frac{1}{4-\xi}ds^2(\text{S}^2_2)+\frac{1}{4}{\cal D}\psi^2\bigg)\bigg].\\[2mm]
				{\cal D}\psi&= d\psi+\cos\theta_1 d\phi_1+\cos\theta_2 d\phi_2+{\cal B},~~~~ {\cal B}=- \frac{2}{\sqrt{N_c}}(1- e^{2(\Phi_0-\Phi)})d\varphi
    \end{aligned}
    \end{equation}
while the RR 3-form flux is still given as in \eqref{eq:3-formaftershift} for 
\beq
p= R_0 e^{2(\Phi_0-\Phi)},~~~R_0= \frac{\sqrt{N_c}}{2},
\eeq
while $\varphi$ is still has period $2\pi R_0$.

This solution interpolates between a regular background at $r=0$ and the UV behaviour of the solution in \eqref{eq:lindil}. In four dimensions, the parameter $\xi$ plays the role of what would be a 'marginal deformation', moving us between different solutions (exploring what would be a 'conformal manifold'). In three dimensions, the role of $\xi$ is similar, exploring a one-parameter family of solutions. It would be good to find some observable that explicitly depends on the parameter $\xi$.

In the next section we will study some field theory aspects of the compactified solutions we have constructed.

\subsection{Some (supersymmetric) cycles}
First we consider the cycle on which the brane sources are extended. Clearly given the form of $\Xi$ this can only be $\Sigma_S=(r,\psi,\varphi)$ evaluate at constant $(\theta_i,\phi_i)$. We find the induced metric and volume from on $\Sigma_S$ are given by
\beq
ds^2(\Sigma_S)=e^{\Phi}\bigg[ \frac{e^{2k}}{\Xi_S}d\varphi^2+  \frac{N_c^2}{e^{2k}}dr^2+ \Xi_S\left(d\psi+{\cal B}\right)^2\bigg],~~~~\Xi_S= \frac{e^{2k}}{4}+p^2, ~~~~\text{vol}(\Sigma_S)= \frac{N_c}{2}e^{\frac{\Phi}{3}}d\Sigma_S.
\eeq
As this is the cycle on which the sources are extended, this must be supersymmetric. We find that
\beq
\Psi^{(7)}_3\bigg\lvert_{\Sigma_S}=\frac{N_c}{2}e^{\frac{\Phi}{3}}d\Sigma_S=\text{vol}(\Sigma_S),
\eeq
so clearly we satisfy \eqref{eq:3-cyclesusy}, which is  the necessary condition for D5 branes extended in Mink$_3$ and wrapping some 3-cycle to be supersymmetric.

Let us study a cycle that  vanishes at the origin (this is useful to define a gauge coupling). Such a cycle is given by $\Sigma_g=(\varphi,\theta_2,\phi_2)$ with $(\theta_1=\theta_2,\phi_2=2\pi-\phi_1)$ and $\psi$ constant. The induced metric and volume form on this cycle are
\beq
\begin{aligned}
ds^2(\Sigma_g) &= e^{\Phi}\bigg[\frac{1}{N_c}(e^{2k}+(\sqrt{N_c}-2 p)^2)d\varphi^2+ (N_c g_0+2 e^{2h}) ds^2(\text{S}^2_2)\bigg],\\ 
\text{vol}(\Sigma_g) &= \sqrt{\frac{1}{N_c}(e^{2k}+(\sqrt{N_c}-2 p)^2)}(N_c g_0+2 e^{2h})d\Sigma_g,
\end{aligned}
\eeq 
this cycle is not supersymmetric, though it need not be. 

We also discuss a cycle that is supersymmetric in the IR which can be used to define a Chern-Simons coupling. We find that $\Sigma_{CS}=(\varphi,\theta_1,\phi_1|\psi=\frac{4}{\sqrt{N_c}}\varphi,r=0)$ for $(\theta_2,\phi_2)$ constant achieves the  goal. We find
\beq\label{eq:SigmaCS}
ds^2(\Sigma_{CS})=\frac{e^{\Phi}_0N_c}{\xi}\Bigg[ds^2(\text{S}^2_1)+\xi\left(\frac{2}{\sqrt{N_c}}d\varphi+ \frac{1}{2}\cos\theta_1 d\phi_1\right)^2\bigg],~~~~\text{vol}(\Sigma_{CS})= \frac{e^{\frac{\Phi_0}{3}}(4 N_c^5)^{\frac{1}{4}}}{\xi}d\Sigma_{CS}.
\eeq
Like $\Sigma_{S}$ this cycle is supersymmetric, but only at $r=0$. Note that
\beq \label{eq:CSInt}
-\frac{1}{(2\pi)^2}\int_{\Sigma_{CS}}F_3= N_c
\eeq
Let us now use all the information gathered in this section to calculate observables of the dual field theory.

\section{Field Theory Observables}\label{sec:Observables}

We proceed to the computation of different observables by using probes of Type IIB, obtaining information about the strong coupling regime of the theory. We first discuss the Chern-Simons term mentioned in \eqref{eq:3DCSQFT}. Then, elaborate on Wilson loops and confinement, finally closing with an indicator of the number of degrees of freedom along the RG-flow.

\subsection{Chern-Simons Level}

We consider a probe D5-brane extended in $(\Mink{3},\varphi,\theta_{1},\phi_{1})$ with $(\psi = \frac{4}{\sqrt{\Nc}}\varphi,r=0)$ and $(\theta_{2},\phi_{2})$ constant, that is a D5-brane wrapping $\Sigma_{CS}$, discussed in \eqref{eq:SigmaCS}. From the DBIWZ action
    \begin{equation}
        S_{DBI-WZ} = T_{D5}\int d^{3}x\sqrt{e^{-2\Phi}\det[g+F]} - T_{D5}  \int_{\Mink{3}\times \Sigma_{CS}} C_{6} + \frac{1}{2}C_{2} \wedge F\wedge F.
    \end{equation}
Here, $F$  is the gauge field on the D5-brane worldvolume (note that the NS-two form is zero in our backgrounds). Focusing on the WZ term and integrating by parts after using \eqref{eq:CSInt} leads to
    \begin{equation}
        \int_{\Mink{3}\times \Sigma_{CS}} C_{2} \wedge F\wedge F 
        = -\int_{\Mink{3}\times \Sigma_{CS}} F_{3} \wedge A \wedge F 
        = \Nc \int_{\Mink{3}}  A \wedge F 
    \end{equation}
We interpret this as the level of the Chern-Simons term generated in the dual theory when integrating out the non-vector-like tower fermionic KK modes of the circle, see the discussion around \eqref{eq:3DCSQFT}. It is interesting to note that this manifold is a calibrated three-cycle only at the end of the space $r=0$.  This is the point in which the geometry captures the ``end'' of the circle compactification, where we have integrated out all the massive fields. At energies higher than the ones set by the size of the circle, it is not possible to integrate out all the massive fields and hence no CS term is generated. This is captured by the dual geometry by the fact that this cycle is calibrated at the point $r = 0$.

\subsection{Wilson Loop}

We consider rectangular Wilson Loops. Following \cite{Maldacena:1998im}, we study the F1 embedding 
    \begin{equation}
        t = \tau, \quad
        x_{1} = \sigma,\quad
        r = r(\sigma).
    \end{equation}
In the class of backgrounds of \eqref{eq:FinalAnsatz}, the induced metric on the F1
    \begin{equation}
        ds^{2}_{\text{Ind}} = e^{\Phi}\left[ -d\tau^{2} + \left( 1 + e^{-2k} \Nc^{2} (r')^{2} \right)d\sigma^{2}  \right],
    \end{equation}
so that the Nambu-Goto action can be written as
    \begin{equation}
       \mathcal{L}_{\text{NG}} = T_{F_{1}} \int d\tau\, d\sigma \sqrt{ F^{2}(r) + G^{2}(r) (r')^{2}}, 
    \end{equation}
with
    \begin{equation}
        F(r) = e^{\Phi}, \quad
        G(r) = \Nc \, e^{\Phi-k}.
    \end{equation}
This form of the Nambu-Goto action matches the one studied in \cite{Nunez:2009da}. We follow the procedure discusses there. The Lagrangian does not depend explicitly on $\sigma$, so the Hamiltonian of the system is conserved. This leads to
    \begin{equation}
        \frac{dr}{d\sigma} = \pm V_{\text{eff}}  ,\quad 
        V_{\text{eff}} = \frac{1}{C}\frac{F(r)}{G(r)}\sqrt{F^2(r)- C^{2}}.
    \end{equation}
where $C= F(r_{0})$, with $r_{0}$ the turning point of the U-shaped embedding.  

The separation and energy are given as functions of $r_0$, the distance from the origin of the radial coordinate, $r=0$, to the position of the turning point of the string. The expressions for these quantities are,
    \begin{align}
        L_{QQ}\left( r_{0}\right) &=2\int_{r_{0}}^{+\infty }
        \frac{dr}{V_{\text{eff}}(r) }\, , \label{QQ separation}  \\
        E_{QQ}\left( r_{0}\right) &=F\left( r_{0}\right) L_{QQ}\left( r_{0}\right)
            +2\int_{r_{0}}^{+\infty }dr\frac{G\left( r\right) }{F\left( r\right) }
            \sqrt{F\left( r\right) ^{2}-F\left( r_{0}\right) ^{2}}-2\int_{0}^{+\infty }dr\
            G\left( r\right) \label{QQ energy} \, .
    \end{align}

One of the conditions of \cite{Nunez:2009da} requires that the effective potential diverges as $V_{\text{eff}} \sim r^{\beta}$ with $\beta>1$ in order to have a finite contribution from the upper end of the integral in \eqref{QQ separation}. In the case of the asymptotically constant dilaton, we can use \eqref{eq:ExpansionUVConstantDilaton} to check that
    \begin{equation}
        \lim_{r \rightarrow +\infty} V_{\text{eff}} \overset{\eqref{eq:ExpansionUVConstantDilaton}}{\sim} \sqrt{r}.
    \end{equation}
Hence, since the length of the loop receives an infinite contribution from the UV, it is not possible to study this loop configuration in this geometry. On the other hand, for the exact solution with linear dilaton in Section \ref{sec:exactsol} we have
    \begin{equation}
        \lim_{r \rightarrow +\infty} V_{\text{eff}} \overset{\eqref{eq:ExactSolutionMetric}}{\sim} e^{r},
    \end{equation}
so that it is possible to study this loop. In fact, it is possible to exactly integrate the length and the energy of the loop. On the solution of Section \ref{sec:exactsol}, we have
    \begin{equation}
        F(r) = e^{r+\Phi_{0}}, \quad
        G(r) = \frac{\sqrt{\Nc} e^{r+\Phi_{0}}}{\sqrt{1-e^{-4r}}}, \quad
        V_{\text{eff}} = \frac{\sqrt{\Nc}}{C} \sqrt{1- e^{-4r}}\sqrt{e^{2r+2\Phi_{0}}-C^{2}}
    \end{equation}
Before computing the length and energy of the loop, it is convenient to perform the change of variables $r = \log(\rho) - \Phi_{0}$. In this variables $C = \rho_{0}$, with $\rho_{0}$ the turning point of the string in the new variable
and $\rho_{+} = e^{\Phi_{0}}$ is the point in which $f(\rho)$ vanishes. The integrals defining the length and the energy of the loop are then mapped to the ones obtained \footnote{Here, our background is supersymmetric, so we have to set $\rho^{2}_{-}=-\rho^{2}_{+}$ when comparing to those references.} in \cite{Nunez:2023nnl,Nunez:2023xgl} when studying rectangular Wilson Loops on similar backgrounds constructed with D5-branes. The reason for this similarity is that the induced metric on the worldsheet is the same.

The integrals have analytic expressions in terms of the Elliptic functions
    \begin{equation}\label{definitionsEK}
        {\bf K}(x)=\int_0^{\frac{\pi}{2}} \frac{d\theta}{\sqrt{1-x \sin^2\theta}}, \quad 
        {\bf E}(x)= \int_0^{\frac{\pi}{2}} \sqrt{1 -x \sin^2\theta} d\theta.
    \end{equation}
Explicitly
    \begin{align}
        L_{QQ} &=2\sqrt{\Nc} \frac{\rho_{0}}{\sqrt{\rho_{0}^{2} -\rho_{+}^{2} }} {\bf K}\left( 
        \frac{2\rho_{+}^{2}}{\rho_{+}^{2}- \rho_{0}^{2}} \right) \,\\
        E_{QQ} &=2 \sqrt{\Nc} \left( \frac{\rho_{0}^{2}}{\sqrt{\rho_{0}^{2}-\rho_{+}^{2}}}\mathbf{K}%
\left( \frac{2\rho_{+}^{2}}{\rho_{+}^{2}-\rho_{0}^{2}}\right)  -\mathbf{E}\left( \frac{2\rho_{+}^{2}}{\rho_{+}^{2}-\rho_{0}^{2}}\right) \sqrt{\rho_{0}^{2}-\rho_{+}^{2}}+\frac{\sqrt{2\pi^{3}}}{\Gamma\left(\frac{1}{4}\right)^{2}} \right)
    \end{align}

Expanding near $\rho_{0}\sim \rho_{+}$ we have
    \begin{align}
        L^{(IR)}_{QQ} &\sim \sqrt{\frac{\Nc}{2}}\log\left( \frac{16 \rho_{+}}{\rho_{0}-\rho_{+}} \right) + \mathcal{O}(\rho_{0}-\rho_{+}),\\
        E^{(IR)}_{QQ} &\sim \sqrt{\frac{\Nc}{2}} \rho_{+} \left( - 4\sqrt{2} + 4\frac{\sqrt{2\pi^{3}}}{\Gamma\left(\frac{1}{4}\right)^{2}} + \sqrt{2}\log\left( \frac{16 \rho_{+}}{\rho_{0}-\rho_{+}} \right) \right) + \mathcal{O}(\rho_{0}-\rho_{+})
    \end{align}
from where we can solve
    \begin{equation}
        \rho_{0} - \rho_{+} \sim  + 16 e^{-\sqrt{\frac{2}{\Nc}}L^{(IR)}_{QQ}} \rho_{+}
    \end{equation}
then
    \begin{equation}
        E^{(IR)}_{QQ} \sim e^{\Phi_{0}} L^{(IR)}_{QQ} +  \frac{\sqrt{\Nc}}{2}e^{\Phi_{0}}\left(  - 4\sqrt{2} + 4\frac{\sqrt{2\pi^{3}}}{\Gamma\left(\frac{1}{4}\right)^{2}}\right) + \mathcal{O}\left(e^{-\sqrt{\frac{2}{\Nc}}L^{(IR)}_{QQ}}\right).
    \end{equation}
Such linear behaviour for large $L$ signals IR confinement.
\subsection{Holographic central charge}

Whilst only well-defined at fixed points, the  notion of central charge can be extended along RG-flows. This is  natural to do in holographic models. We briefly discuss a quantity that provides a measure of the number of degrees of freedom in our system.  Since the field theory is anisotropic (due to the circle compactification), we should use a definition for holographic central charge given in \cite{Bea:2015fja}, see \cite{Merrikin:2022yho}
for a  summary and generalisation. Comparing our background
in (\ref{eq:FinalAnsatz}) with the generic anisotropic backgrounds discussed in \cite{Bea:2015fja, Merrikin:2022yho} we calculate
\begin{eqnarray}
c_{\text{flow}}=\frac{27 {\cal N}}{8 G_{N,10}}\frac{ e^{2\Phi+ 2h+2g + k+3\tilde{k}}(1+4p^2 e^{-2k})}{\left( 2\Phi' +2h'+2g'+k'\right)^3}, \quad
{\cal N}= 8\pi^3 L_{x_1}L_{x_2}L_\varphi \label{cflowfinal}
\end{eqnarray}
We analyse the central charge near the end of the space using \eqref{eq:IRseries}. We find
    \begin{equation}
        c_{\text{flow}} \underset{r\rightarrow 0}{\sim}  r^{\frac{3}{2}}  + \mathcal{O}\left(r^{\frac{5}{2}}\right), 
    \end{equation}
which indicates that the number of degrees of freedom goes to zero at the end of the flow, the QFT is gapped. This is consistent with the fact that 3D $\calN=2$ SU($\Nc$)$_{\Nc}$ is a gapped  theory. This behaviour is universal for all our solutions as they share the same IR.

On the other hand, we compare the two different UV behaviours of the two solutions  in the previous section. We find,
    \begin{equation}
        c_{\text{flow}} \underset{r\rightarrow +\infty}{\sim} 
        \begin{cases}
            r^{4} + \mathcal{O}\left(r^{3}\right), \quad \text{Asymptotically constant dilaton} \eqref{eq:ExpansionUVConstantDilaton},\\
            e^{2r} + \mathcal{O}\left(e^{-2r}\right), \quad
            \text{Linear dilaton}. \eqref{eq:ExactSolutionMetric}
        \end{cases}
    \end{equation}
The number of degrees of freedom grows unbounded for high energies. This is not unexpected, our QFTs are not completed by a UV-fixed point. On the contrary, the UV completion is in terms of a Little String Theory.

\section{Conclusion and Open Questions}

Let us conclude this paper with a concise summary of its key results and a discussion of potential directions for future research.

This work focuses on the holographic duals of a version of four-dimensional ${\cal N}=1$ SU$(\Nc)$ SQCD in the special case where the number of flavours  and colours satisfy the relation  $N_f = 2N_c$. In this case, the R-symmetry is non-anomalous. The type IIB supergravity backgrounds dual to this theory were originally constructed in \cite{Casero:2006pt, Casero:2007jj, Hoyos-Badajoz:2008znk}, and serve as the ‘seed’ for the new solutions developed in this paper.

We implement a twisted-circle reduction of the SQFT, which crucially requires the presence of a conserved R-symmetry. The holographic construction follows the formalism laid out in \cite{Macpherson:2024qfi}. Within this framework, we uncover novel holographic backgrounds whose {\it regularity requires} the condition $N_f = 2N_c$. This outcome aligns perfectly with the requirement of an anomaly-free R-symmetry in the original theory. It is both compelling and insightful to observe how the gravitational dual encodes the anomaly cancellation condition for the twisted compactification directly in the solutions of the BPS equations.

Looking ahead, several promising avenues merit further exploration:
\begin{itemize}
\item{Investigating whether our logic admits any limitations. Specifically, is it possible to perform a consistent twist using an anomalous R-symmetry, or perhaps another global symmetry? Affirmative results in this direction could potentially yield smooth backgrounds for theories with $N_f \neq 2N_c$.}
\item Extend the results of \cite{Cassani:2021fyv} to include more that one gauge group and bifundamental matter. The supergravity solutions of \cite{Chatzis:2024kdu,Chatzis:2024top,Castellani:2024ial}  suggest that a similar construction for quiver gauge theories should be possible.
\item {Exploring whether analogous constructions can be extended to six-dimensional SQFTs. As in three-dimensions, integrating out a fermion in a five-dimensional theory also generates a Chern-Simons term depending on the sign of the mass of the fermion. Then, it would be interesting to check whether performing a twisted circle compactification of a given 6D theory to 5D, also leads to a KK spectrum of hypermultiplets which is non vector-like, and that when integrated out, the generated Chern-Simons level is precisely $\Nc$.}  
\item{It would be nice to study the addition of a black hole on top of our solution. }
\item Study similar mechanisms from theories in odd dimensions compactified to even dimensions. In these cases there is no Chern-Simons term generated in the effective lower dimensional theory (since it is even dimensional). It would be interesting to see what is the even dimensional counterpart of that. 
\end{itemize}

We aim to address these and related questions in future work.

\section*{Acknowledgements}

We thank Riccardo Argurio, Matteo Bertolini, Francesco Bigazzi, Nikolay Bobev, Davide Cassani, Aldo Cotrone, S. Prem Kumar, Francesco Mignosa, Angel Paredes and Alfonso V. Ramallo, who shared their knowledge with us. The work of N. M. and R. S. is supported by the Ram\'on y Cajal fellowship RYC2021-033794-I, and by grants from the Spanish government MCIU-22-PID2021-123021NB-I00 and principality of Asturias SV-PA-21-AYUD/2021/52177. P. M. and  C. N. are supported by the grants ST/Y509644-1, ST/X000648/1 and ST/T000813/1.


\appendix

\bibliographystyle{JHEP}

\bibliography{refs}

\end{document}